\documentstyle[aps,prl,multicol,epsf]{revtex}

\def\K{G}
\def\Ylm{\mathrm{Y}_{lm}}

\begin{document}

\draft
\title{Hexatic Order and Surface Ripples in Spherical Geometries}
\author{Peter Lenz and David R. Nelson}
\address{Lyman Laboratory of Physics, Harvard
  University, Cambridge, MA 02138}
\date{\today }
\maketitle

\begin{abstract}
In flat geometries, two dimensional hexatic order has only a minor
effect on capillary waves on a liquid substrate and on undulation
modes in lipid bilayers. However, extended bond orientational order
alters the long wavelength spectrum of these ripples in spherical
geometries. We calculate this frequency shift and suggest that it
might be detectable in lipid bilayer vesicles, 
at the surface of liquid metals and in multielectron bubbles in liquid 
helium at low temperatures. Hexatic order also leads to a shift in the 
threshold for  the fission
instability induced in the later two systems by an excess of electric
charge. 
\end{abstract}


\begin{multicols}{2}
\narrowtext

One of the main predictions of theories of dislocation
\cite{kost73,halp78} and disclination \cite{halp78} mediated melting
in two dimensions is that the transition from solid to liquid can be a
two-stage process \cite{halp78}.  First, at a temperature $T=T_m$,
dissociation of dislocation pairs drives the transition from a crystal
to an intervening hexatic phase.  This liquid-crystalline phase has no
translational order but still exhibits extended orientational
correlations. Then, its quasi-long-ranged orientational order is
destroyed in a second transition which takes place at a higher
temperature $T=T_i$. Here, an unbinding of disclination pairs finally
produces an isotropic liquid.

The hexatic phase predicted by theory has now been observed in free
standing liquid crystal films \cite{chou98}, in various types of two
dimensional colloidal crystals \cite{murr92,zahn99}, in magnetic
bubble arrays \cite{sesh91} and in Langmuir-Blodgett surfactant
monolayers \cite{knob92}. Despite the modest time scales available
even on the fastest computers, there is now evidence via computer
simulations for continuous melting and a narrow sliver of hexatic
phase for hard disks \cite{jast99} and a more substantial hexatic
phase for particles interacting with a repulsive $1/r^{12}$ potential
\cite{bagc96}.

In the above experiments, two dimensional order was typically probed
via diffraction or by direct measurement of correlation functions in
real space. It is difficult to use these methods when hexatic order is
present in a curved, spherical geometry. Examples where hexatic order
might be present include "liposomes", i.e., closed vesicles composed
of lipid bilayers \cite{park92}, the surface of liquid metal droplets
confined in Paul traps \cite{davi97}, and multielectron bubbles
submerged in liquid helium \cite{leid95}. Two dimensional planar
multilayers of the phospholipid DMPC \cite{chia95} have many
similarities to free standing liquid crystal films, 
which are known to
have hexatic phases \cite{chou98}. Celestini et al. have found
evidence from computer simulations for extended orientational
correlations at the surface of supercooled liquid metals
\cite{cele97}. Electrons trapped on the surface of liquid helium by a
submerged, positively charged capacitor plate have long 
been used to
investigate two dimensional melting \cite{grim79}. Although these
experiments are {\em consistent} with dislocation mediated melting,
evidence for a hexatic phase and disclination unbinding remains
elusive. Multielectron bubbles result when large numbers of electrons
$(10^5-10^7)$ at the helium interface subduct in response to an
increase in the anode potential and coat the inside wall of a large
sphere (10-100 micron radius) of helium vapor.

One might hope that bond orientational order in a membrane or
interface could be detected by its effects on the dynamics of
undulation modes or capillary waves. Unfortunately, hexatic order
couples only to the {\em Gaussian} curvature \cite{nels87}, which
vanishes for a simple sine wave deformation of a flat membrane or an
interface (cf. Eq.~(\ref{eq:n11}) below). The situation is different,
however, when these excitations are superimposed on a nontrivial
background geometry such as that of a sphere. Here, we determine the
effect of hexatic order on, e.g., the quadrupole and octopole
undulation modes and capillary wave excitations for the spherical
systems described above. The frequency shift is large for liposomes
with hexatic order and might also be observable in liquid metal
droplets and in multielectron bubbles in helium.

We first consider liquid droplets with surface hexatic order, 
either due to surface ordering or
to coating with surface active molecules. The
equilibrium shape minimizes a free energy $F_d$ given by
\begin{equation}
\label{eq:n1}
    F_d =
\sigma \int dA
 + \frac{1}{2} K_A \int dA  \ D _in^j D ^i n_j,
\end{equation}
where $\sigma$ denotes the surface tension of the interface connecting
a droplet with density $\rho_l$ surrounded by vapor with density
$\rho_v$.  For a general manifold with internal coordinates
$x=(x_1,x_2)$, the surface element is given by $dA=\sqrt{g} d^2x$,
where $g(x)$ is the determinant of the metric tensor $g_{ij}(x)$. For
an undeformed sphere with radius $R_0$, $x \equiv (\theta,\varphi)$
with polar coordinates $\theta$ and $\varphi$ and 
$dA=R_0^2 \sin\theta d\theta d \varphi$.  
Finally, $\vec{n}$ is a unit vector in the
tangent plane with $n_i n^i=1$ which identifies (modulo $2 \pi/6$) the
long-range correlations in the hexatic bond directions \cite{nels87}.
Here, $D_i n^j \equiv g^{jk}D_in_k$, where $g^{ij}$ is the inverse of
$g_{ij}$ and $D_i$ denotes a covariant derivative with respect to the
metric $g_{ij}$. Thus, 
$D_in^j \equiv \partial _i n^j+\Gamma_{ki}^jn^k$, 
where the $\Gamma_{ki}^j$ are Christoffel symbols of second kind,
see e.g. \cite{krey}.  The hexatic stiffness 
$K_A \sim E_c (\xi_T/a_0)^2$, 
where $\xi_T$ is the translational correlation length,
$a_0$ is the particle spacing and $E_c$ is the dislocation core
energy. The ratio $K_A/k_BT$ jumps from an universal value $72/\pi$ to
zero when the hexatic melts into an isotropic liquid at $T=T_i^-$
\cite{halp78}.

The fundamental assumption which underlies the hexatic free energy
discussed above is that the configuration of minimal elastic energy
corresponds to a vector field $\vec{n}_0$ where $\vec{n}_0(x+dx)$ can
be obtained from $\vec{n}_0(x)$ by parallel transport of $\vec{n}_0$.
On a sphere however, curvature introduces ``frustration'' since parallel
transport of $\vec{n}$ along closed loops on the surface leads to a
rotation of $\vec{n}$.  For geometries frustrated by a nonzero
integrated Gaussian curvature $\overline{\K}$, the state of minimum
energy always has topological defects \cite{nels83}.

These defects reduce the elastic energy by screening the Gaussian
curvature.  This point can be made more precise by introducing a local
bond-angle field $\theta$, the angle between $\vec{n}$ and some local
reference frame. The singular part of $\theta$ is then connected with
the disclination density \cite{nels87,nels83} and the hexatic
contribution $F_h$ to Eq.~(\ref{eq:n1}) becomes \cite{bowi00}
\begin{eqnarray}
    F_h  & = &  -\frac{1}{2} K_A \int dA \int dA'
 \left [\K(x)-s(x) \right] \nonumber \\
& & \times \left (\frac{1}{\Delta_{xx'}} \right )
 \left [\K(x')-s(x') \right] . \label{eq:2}
\end{eqnarray}
Here, $1/\Delta$ is the inverse Laplacian, $\K(x)$ the
Gaussian curvature and $s(x)$  the 
disclination density \cite{nels83}, 
\begin{equation}
\label{eq:3}
s(x) \equiv 
\frac{1}{ \sqrt{g(x)}} \sum_{i=1}^{N_d} q_i \delta(x-x_i),
\end{equation}
with $N_d$ disclinations of charge $q_i=\pm 2 \pi/6$ at positions
$x_i$.  The defects minimize $F_h$ by arranging themselves to
approximately match the Gaussian curvature. For low temperatures,
large core energy $E_c$ and spherical geometries we expect $N_d=12$,
corresponding to 12 \ 5-fold disclinations at the vertices of an
icosahedron.

To investigate the influence of hexatic order on droplets, we
study deformations about the equilibrium configuration. We expand the
free energy $F$ in a small time-dependent displacement field 
$\delta\vec{R}(x,t)$, where $\vec{R}'=\vec{R}_0+\delta \vec{R}$ 
is the
deformed surface and $\vec{R}_0$ is the radius vector of a sphere.
Here, the displacement field can be chosen to be purely normal, 
$\delta \vec{R}(x,t)=R_0 \zeta(x,t) \vec{N}$,  
where $\vec{N}=\vec{R}_0/R_0$ is
the normal vector of the sphere and $\zeta$ can be expanded in terms
of spherical harmonics
\begin{equation}
\zeta(x,t)=\sum_{l=0}^{\infty } \sum_{m=-l}^{l} r_{lm}(t)\Ylm(x).
\end{equation}
In the absence of defects, the expansion of  $F$ in $\zeta$ would be
straightforward. On the sphere however,
one has to deal with a distribution of discrete disclination charges
which produces a small static icosahedral surface deformation. We 
initially neglect this discreteness and show afterwards that the 
corrections arising from the discrete nature of $s(x)$ 
are irrelevant  for the 
oscillation frequencies $\omega(l)$ with $l<6$.
These considerations can be made more precise by expanding $s(x)$ in
terms of 
spherical harmonics 
\begin{equation}
\label{eq:5}
  s(x)=\K_0+\frac{1}{R_0^2} \sum_{l=1}^{\infty }\sum_{m=-l}^{l}
  s_{lm} \Ylm(x) ,  
\end{equation}
where $\K_0=1/R_0^2$ and 
$s_{lm} \equiv 
\sum_{i=1}^{N_d}q_i{\mathrm{Y}_{lm}^*}(x_i)$.
In the following, we 
first assume 
$s_{lm}\simeq 0$ for $l > 0$ and then discuss the 
corrections due to nonzero
$s_{lm}$.

For $s_{lm}=0$ with $l>0$ 
the hexatic order has no influence on the droplet shape
and the equilibrium configuration is a sphere with a radius $R_0$.
However, the presence of hexatic order and a nonzero stiffness
constant $K_A$ does affect the fluctuation spectrum $\omega(l)$ of a
hexatic liquid droplet. To see this, we solve the Navier-Stokes
equation for the fluid in the presence of an undulating interface with
boundary condition \cite{land2}
\begin{equation}
\label{eq:n6}
  \rho_l \left. \frac{\partial \Phi (r)}{\partial t} 
  \right |_{r=(R_0+\zeta R_0)^-} 
-  \rho_v \left. \frac{\partial \Phi (r)}{\partial t} 
  \right |_{r=(R_0+\zeta R_0)^+} 
= \Delta p (x),
\end{equation}
where $\Phi(r,x,t)$ is the velocity
potential. 
Eq.~(\ref{eq:n6}) thus relates the 
pressure difference between the inside and outside of the droplet
with the 
generalized pressure discontinuity $\Delta p$ which is caused by the
shape displacement. 
$\Phi(r,x,t)$ is here given by \cite{land2}
$\Phi(r,x,t) =\sum_{l,m} A_{lm}^> \Ylm(x) \left ( \frac{R_0}{r}\right
)^{l+1}$ for $r>R_0(1+\zeta)$ and 
$\Phi(r,x,t)=\sum_{l,m} A_{lm}^< \Ylm(x) \left ( \frac{r}{R_0}\right
)^{l}$ for $ r<R_0(1+\zeta)$. 
The displacement field $\zeta$ and the velocity potential $\Phi$ are
related by the additional boundary condition
$R_0 \dot{\zeta} \equiv R_0\partial \zeta/\partial t=
\left. \partial \Phi/\partial  r\right |_{r=R_0(1+\zeta)}$, which yields 
$A_{lm}^>(t)=-R_0^2 \dot{r}_{lm}(t)/(l+1)$
and $A_{lm}^<(t)=R_0^2 \dot{r}_{lm}(t)/l$. Upon setting
$\Delta p(x) \equiv \sum_{l,m} (\Delta p_{lm}(t))\Ylm(x)$, 
one then finds
\begin{equation}
\Delta p_{lm}(t)=\left (
    \frac{\rho_l}{l}+\frac{\rho_v}{l+1}\right ) R_0^2 \ddot{r}_{lm}(t),
\end{equation}
where the generalized surface pressure of the displaced surface is
given by 
$\Delta p_{lm}=- \delta F(r_{lm})/R_0^3 \delta r_{lm}^*$. 
Since the interfacial energy contribution to Eq.~(\ref{eq:n1}) is 
(cf. e.g. \cite{land2})
\begin{equation}
 F_{i}=\frac{1}{2}\sigma R_0^2\sum_{l,m}|r_{lm}|^2(l-1)(l+2) 
\end{equation}
and the hexatic free energy reads \cite{lenz00a}
\begin{equation}
 F_h=\frac{1}{2}K_A   \sum_{l=1}^{\infty } \sum_{m=-l}^{l}
|r_{lm}|^2
  \frac{(l-1)^2(l+2)^2}{l(l+1)} 
\end{equation}
we set $r_{lm}=r^0_{lm}(t)e^{-i \omega(l)t}$ and find
(for $l>0$ and $\rho_{v} \ll \rho_{l}$)
\begin{equation}
\label{eq:n10a}
  \omega^2=\frac{\sigma}{\rho_{l}R_0^3}l(l-1)(l+2)\left [
    1+\frac{K_A}{\sigma R_0^2} \frac{(l-1)(l+2)}{l(l+1)} \right ].
\end{equation}                                            

Note that 
$\omega(l)$ vanishes for $l=1$,  corresponding to 
translations of the droplet as a whole. 
Eq.~(\ref{eq:n10a}) also shows that hexatic order only 
affects capillary waves in a curved geometry: In
the flat space limit of large 
$R_0$ and $l \gg 1$ with 
$k \equiv l/R_0$ fixed, one has
\begin{equation}
\label{eq:n11}
  \omega^2 \simeq \frac{k^3}{\rho_l} \left [ \sigma+
    \frac{K_A}{R_0^2}\right ] .
\end{equation}
The hexatic contribution drops out as $R_0 \rightarrow \infty$ and we
recover the result for capillary waves of a flat fluid
surface \cite{land2}.  Thus, it is essential to study deformations of
a curved geometry  to reveal the
presence of hexatic order.  In general, the undulation frequency
Eq.~(\ref{eq:n10a}) depends on the ratio $\frac{K_A}{\sigma R_0^2}$
which for hexatic order at the surface of supercooled liquid metal
droplets is $K_A/\sigma R_0^2 \simeq (\xi_T/R_0)^2(E_c/\sigma a_0^2)$.
This ratio becomes of order unity when $\xi_T \approx R_0$.

\begin{figure}
  \begin{center}
    \mbox{\epsfxsize=5cm  \epsfclipon     \epsffile{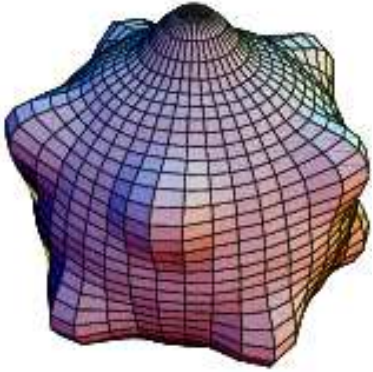}}
  \end{center}
\caption{\label{fig2}
Shape of a liquid droplet with hexatic order and 12 disclinations
lying on the vertices of an
icosahedron (surface deformations associated with the defects 
are exaggerated).}
\end{figure}

Nonzero coefficients $s_{lm}$ with $l>0$ affect the mean curvature 
$H(x)$ of the stationary droplet via the extremal equation
\cite{lenz00a} 
\[
  2H=2H_0+\frac{K_A}{\sigma R^2_0} \frac{1}{R_0}
  \sum_{l=1}^{\infty } \sum_{m=-l}^{l}
  \frac{(l-1)(l+2)}{l(l+1)}s_{lm}\Ylm  , 
\]
with $H_0=1/R_0$, leading to 
static surface deformation coefficients 
$r_{lm}^0=s_{lm}K_A/\sigma R_0^2l(l+1)$ which {\em vanish} in the limit 
$R_0 \rightarrow \infty$.
Thus, for $K_A \neq 0$ the defects effectively repel each other and
deform the droplet, cf. Fig.~\ref{fig2}. 
However, as we show below, nonzero $s_{lm}$ have no influence 
on the frequencies $\omega(l)$ for $0<l<6$.

Next, we discuss the influence of hexatic order on the fluctuations of
a vesicle. Here, the free energy is given by \cite{park92,nels87}
\begin{eqnarray}
    F_v & = & \frac{1}{2} \kappa \int dA \  (2H)^2+\kappa_G\int dA \ \K
\nonumber \\
& &  + \frac{1}{2} K_A \int dA  \ D _in^j D ^i n_j, 
\label{eq:n13}
\end{eqnarray}
where $\kappa$ and $\kappa_G$ are the mean and Gaussian rigidity,
respectively. In the following analysis, the second term of
Eq.~(\ref{eq:n13}) can be neglected since we only consider surface
shapes which are topologically equivalent to a sphere.

Since $\kappa$ plays a similar role for vesicles as $\sigma$ plays for
droplets, hexatic order should lead here to similar effects on the
spectrum $\omega(l)$.  However, for vesicles the fluctuations are
overdamped and one has to analyze the Stokes equation. This can be
done by generalizing the approach of \cite{schn84} to hexatic
membranes.  One then obtains \cite{lenz00a} (neglecting  
for simplicity the volume constraint)
\begin{eqnarray}
  \omega(l) & = & -i\frac{\Gamma(l)}{\eta R_0^3 } (l-1)(l+2)
\nonumber \\
& & \times \left [\kappa 
  l(l+1)+ K_A \frac{(l-1)(l+2)}{l(l+1)}\right ],
\label{eq:n10}
\end{eqnarray}
where $\eta$ is the liquid viscosity and
$\Gamma(l) \equiv l(l+1)/(2l+1)(2l^2+2l-1)$. In the flat space limit 
one has
\begin{equation}
  \omega \simeq -i \frac{1}{4\eta} \left [ \kappa k^3+
    \frac{K_A}{R_0^2}k\right ] ,
\end{equation}
in agreement with \cite{seif97} for $R_0 \rightarrow \infty$.

The frequency shift (\ref{eq:n10}) now depends on the ratio
$K_A/\kappa$. However, as $R_0 \rightarrow \infty$, we expect that
$K_A \simeq 4\kappa$ (a {\em universal} result for flat hexatic
membranes at long wavelength \cite{davi87}) leading to a frequency
enhancement by a factor $\simeq 13/9 \simeq 1.44$ for the $l=2$
quadrupole mode.  Thus, bond orientation order has a strong effect on
the fluctuations of a membrane and should have experimentally
observable consequences in dynamical light scattering \cite{schn84}.

For vesicles, the presence of a finite number of defects also leads to
an equilibrium configuration with a deformed surface. The mean
curvature $H(x)$ of the stationary vesicle now satisfies
\[
  2H=2H_0+\frac{K_A}{\kappa} \frac{1}{R_0}
  \sum_{l=1}^{\infty } \sum_{m=-l}^{l}
  \frac{(l-1)(l+2)}{l^2(l+1)^2}s_{lm}\Ylm  , 
\]
leading to nonzero coefficients 
$r_{lm}^0=s_{lm}K_A/\kappa l^2(l+1)^2$ in the
ground state. However, icosahedral symmetry insures that $s_{lm}=0$
unless 
$l=6,10,12,...$, so corrections of order $s_{lm}$ have no influence on
the frequencies $\omega(l)$ for small $l$. Provided the positions of
the disclinations remain fixed in the lipid matrix on the time scale
of an undulation ($\omega(l=2) \simeq 40$Hz for a $1 \mu$ vesicle) the
dispersion relation (\ref{eq:n10}) remains valid for $0<l<6$.  Because
disclination motion is catalyzed by absorption and emission of
dislocations with spacing $\xi_T$, 
the disclination diffusion constant is 
$D_5 \approx (a_0/\xi_T)^2D_{lipid}$, 
where $D_{lipid} \simeq 10^{-8}\mathrm{cm}^{2}/\mathrm{sec}$. 
The estimate $(a_0/\xi_T)^2 \simeq 10^{-2}$ suggests only
minor disclination motion during an undulation period.  Similar
estimates show that disclination motion is negligible during a
capillary wave period for hexatic droplets.

Finally, we will discuss 
multielectron bubbles in liquid $ ^4\mathrm{He}$. These bubbles can
undergo both a freezing 
transition and a shape instability. Thus, here hexatic order 
affects not
only  the fluctuation spectrum but also the
instability-threshold for fission.
The free energy of a multielectron bubble $F_b=F_d+F_c$ is that of a
droplet (cf. Eq.~(\ref{eq:n1})) with an additional Coulomb
contribution 
\begin{equation}
 F_c=\frac{1}{2 \varepsilon} \int dA \int  dA' \
 \frac{\rho(x)\rho(x')}{|x-x'|},  
\end{equation}
where $\rho(x)$ denotes the charge distribution on the surface and
$\varepsilon$ is the $ ^4\mathrm{He}$ dielectric constant.
In an equilibrium fluid, $\rho=eN/4 \pi R_0^2$ for a sphere with $N$
electrons.

Within the approximations described above, one now finds
(with $l>0$ and neglecting the density {\em inside} 
the bubble) \cite{lenz00a}
\begin{eqnarray}
  \omega^2 & = & \frac{\sigma}{\rho_{l}R_0^3}(l-1)(l+1)
\nonumber \\
& & \times \left [
    (l+2)-4 \frac{R_{cr}^3}{R_0^3}+\frac{K_A}{\sigma R_0^2}
    \frac{(l-1)(l+2)^2}{l(l+1)} \right ],
\end{eqnarray}              
where $R_{cr}^3=(eN)^2/16 \pi \sigma \varepsilon$ is the critical
radius for  
multielectron bubbles without hexatic order \cite{salo81}. 
Thus, for $K_A=0$
spherical bubbles become unstable to fission if $R<R_{cr}$, i.e. 
$\omega^2(l=l_c)<0$ for $R<R_{cr}$ and $l_c=2$. For $K_A \neq 0$ the
stability of charged bubbles is enhanced 
by the hexatic order of the electrons on the
sphere.  Thus, for $T_m<T<T_i$ one still has $l_c=2$ but
$\omega^2(l_c=2)<0$ for $R_0<R_c'$ with $R_{c}'<R_{c}$.
The icosahedral 
symmetry of the deformed shape with  $s_{lm}\neq 0$ is too high to
have an influence on the fission instability which occurs at $l=2$.

Because the electrons (which determine $K_A$) are far apart relative
to the helium atoms (which determine $\sigma$) $K_A/\sigma R_c^2$ will
be smaller than for droplets of supercooled liquid metals with 
$R_0 \simeq R_c$.  
For helium-bubbles with $N \simeq 10^6$ one has
$R_c \simeq 10 \mu$m and we expect that 
$\frac{K_A}{\sigma R_c^2}\simeq 10^{-3}$.  
{\em Charged} metal droplets undergo the same
fission instability.  Thus, it might be possible to detect hexatic
order by investigating the stability of charged liquid droplets in
Paul traps.

The effects discussed here are even larger if the hexatic phase is
bypassed and one freezes directly into a 
curved two-dimensional solid with
shear modulus $\mu$ \cite{zhan93}. The resulting frequency shifts can 
be estimated by
setting $K_A \simeq \mu R_0^2$ in the formulas above (although the
$l$-dependence will be different). There are 
also interesting consequences of hexatic order for 
dynamics of liquids and 
membranes in a cylindrical geometry \cite{lenz00a}.

We thank S. Balibar, G. Gabrielse, N. Goddard, J. Lidmar and
R. Pindak for helpful
discussions.  This work has been supported by the NSF through Grant
No.~DMR97-14725 and through the Harvard MRSEC via Grant
No.~DMR98-09363. P.~L. has been supported by an Emmy-Noether
fellowship of the Deutsche
Forschungsgemeinschaft (Le 1214/1-1).

\end{multicols}
\end{document}